\documentclass[doublecol]{epl2} 

\usepackage{epsfig,epstopdf}
\usepackage{bm,epstopdf}
\usepackage{graphicx,times}
\usepackage{amssymb,amsmath,amsfonts,amsthm}

\newcommand{\ket}[1]{|{#1}\rangle}
\newcommand{\bra}[1]{\langle{#1}|}

\newcommand{\exval}[1]{\langle{#1}\rangle}

\title{Parametric description of the quantum measurement process}

\author{P. Liuzzo-Scorpo\inst{1,2} \and A. Cuccoli\inst{2,3} \and P. Verrucchi\inst{4,2,3}}

\institute{                    
  \inst{1} School of Mathematical Sciences, 
The University of Nottingham, University Park, Nottingham NG7 2RD (UK)\\
  \inst{2} Dipartimento di Fisica ed Astronomia dell'Universit\`a di 
Firenze. Via G.~Sansone 1, I-50019 Sesto F.no, Firenze (Italy)\\
  \inst{3} Istituto nazionale di Fisica Nucleare, sezione di Firenze.
Via G.~Sansone 1, I-50019 Sesto F.no, Firenze (Italy)\\
  \inst{4} Istituto dei Sistemi Complessi ISC-CNR, UoS Dipartimento di 
Fisica. Via G.~Sansone 1, I-50019 Sesto F.no, Firenze (Italy)}
\pacs{}{}

\abstract{
We present a description of the measurement process based 
on the parametric representation with environmental coherent states. 
This representation is specifically tailored for studying
quantum systems whose environment needs being 
considered through the quantum-to-classical cross-over. Focusing upon 
projective measures, and exploiting the
connection between large-$N$ quantum theories and the classical limit of 
related ones, we manage to push our description beyond the pre-measurement step.
This allows us to show that the outcome production follows from 
a global-symmetry breaking, entailing the observed system's state 
reduction, and that the statistical nature of the process is brought 
about, together 
with the Born's rule, by the macroscopic character of the 
measuring apparatus.}

\begin{document}

\maketitle

\section{Introduction} 
\label{S.Intro} 

The measurement process plays a fundamental role in any physical theory, 
and indeed a crucial one in quantum mechanics. The set of works 
dedicated to the subject is almost uncountable and different 
descriptions of the process often implies
different interpretation of quantum mechanics itself (see for instance
Refs.~\cite{BuschLM96,Schlosshauer07} for comprehensive discussions and 
thorough bibliographies). The idea that measuring be a dynamical process
is quite intuitive, and generally accepted. Moreover, it is today 
recognized that such process must 
include an early stage during which entanglement is generated between the 
observed system and the measuring apparatus,
which dictates a quantum treatment for both systems.
This initial stage, usually referred to as 
pre-measurement, can be 
considered successfully completed when the apparatus is in a state that 
conveys 
information on the observed system, a condition related with the 
occurrence of decoherence\cite{Zurek03,Schlosshauer07}.
We usually take what happens next from the measurement postulate, 
although whether or not its content can be derived from the other 
postulates is still the most debated issue of quantum theory
\cite{vonNeumann96,GhirardiRW86,BuschLM96,NamikiPN97,Mermin98,Zurek03,Schlosshauer07,HeinosaariZ12}.

In this work, embodying the
formal definition of the large-$N$ limit of a quantum 
theory\cite{Yaffe82} into the parametric representation with 
environmental coherent states 
(PRECS)\cite{CCGV13pnas,CCGV13osid,Calvani13phd},
we complement the description of the 
measurement process with the quantum-to-classical cross-over 
of the apparatus only. We thus manage to understand how
objective results emerge from the entangled states of a 
macroscopic measuring apparatus, and why their production 
is an inherently statistical phenomenon. The 
observed system's state reduction and the Born's rule are 
contextually obtained.

The structure of the paper is as follows: We first introduce the 
standard model for unitary pre-measurements 
\cite{vonNeumann96,BuschLM96,HeinosaariZ12} and study its dynamics by 
the PRECS. The condition that defines the successful completion of the 
pre-measurement is then formally expressed\cite{L-SCV15ijtp}, and 
two paradigmatic models (spin-$\frac{1}{2}$ in either a bosonic or a 
magnetic environment) are introduced, in order to clarify the formalism 
and guide the interpretation.
At this point we formally associate the classical limit of the 
environment to the macroscopic character of the measuring apparatus 
referring to the seminal work by L.G.Yaffe\cite{Yaffe82} on large-$N$ 
quantum theories;
this finally allows us to describe the outcome production.
In order to ease the understanding of our proposal, we keep comments to
a minimum throughout its presentation, and postpone them to the 
concluding section. 

\section{Pre-measurement and parametric representation}

Let us first outline the formalism to which we will essentially 
refer in describing the quantum measurement process.
Be $\Gamma$ the observed quantum system (with Hilbert space 
${\cal H}_\Gamma$), and $\Xi$ the measuring apparatus (with Hilbert space 
${\cal H}_\Xi$). The composite system $\Psi{{=}}\Gamma{+}\Xi$ is assumed
isolated, i.e. $\rho_\Psi(t){{=}}\ket{\Psi(t)}\bra{\Psi(t)}$, at any time
prior to the output production ($t{<}T$); 
moreover its state is taken separable before the measurement 
starts ($t{\leq}0$)
\begin{equation}
\ket{\Psi(t\le 0)}=\ket{\Gamma}\otimes\ket{\Xi}~.
\label{e.Psi0}
\end{equation}
Note that the validity of these assumptions should not be taken for 
granted, as extensively discussed in Refs.\cite{BuschLM96,Schlosshauer07}.

Let us concentrate upon {\it sharp} observables,  defined as - and 
identified with - projection operator valued measures on the real Borel
space, or a subset of it\cite{BuschLM96}. Any
such measure, $O_\Gamma$, determines a unique Hermitian operator
$\hat O_\Gamma$ acting on ${\cal H}_\Gamma$;
if the observable is further assumed (for the sake of clarity) discrete 
and non-degenerate, it is
$\hat{O}_\Gamma{=}\sum_{\gamma}\omega_\gamma\ket{\gamma}\bra{\gamma}$,
and the $\hat O_\Gamma$-eigenvectors $\{\ket{\gamma}\}_{{\cal 
H}_\Gamma}$ form an orthonormal basis for ${\cal H}_\Gamma$. 
Further ingredients of a scheme designed for describing the measure of 
$O_\Gamma$ are {\it i)} a pointer observable $O_\Xi$ of $\Xi$, {\it ii)} 
a pointer function correlating the 
value sets of $O_\Xi$ and $O_\Gamma$, {\it iii)} a measurement coupling
between $\Gamma$ and $\Xi$, ultimately responsible for the $\Psi$-state 
transformation $\rho_\Psi(0)\xrightarrow{V}\rho_\Psi(t)$
occurring before the actual production of a specific output is obtained.
It can be shown that a sufficient condition for a state transformation 
to qualify as a proper pre-measurement\cite{BuschLM96}, is 
that $V$ be a trace-preserving linear mapping. When $V$ is further 
assumed to be unitary, the process coincides with the one first 
described by von Neumann\cite{vonNeumann96}, later generalized by several 
authors
\cite{Wigner52,ArakiY60,Yanase61,ShimonyS79}, and characterized by 
Ozawa\cite{Ozawa84} under the name of {\it conventional measuring 
process}. If $O_\Xi$ is a sharp observable, with $\hat O_\Xi$ the corresponding 
hermitian operator, choosing $V{=}e^{-it\hat{H}_\Psi}$ 
with 
\begin{equation}
\hat{H}_\Psi=u\,\hat{O_\Gamma}\otimes\hat{O_\Xi}
+\hat{{\rm I\!I}}_{\Gamma}\otimes\hat{H}_\Xi~,
\label{e.H}
\end{equation}
defines the {\it standard model}\cite{BuschLM96} 
for describing pre-measurements as dynamical processes,
where $\hat{H}_\Xi$ acts on 
${\cal H}_\Xi$, $\hat{{\rm I\!I}}_\Gamma$ is the identity operator on 
${\cal H}_\Gamma$, and we have set $\hbar{=}1$.
In what follows we will specifically study the above standard 
model, taking the identity as the pointer function, for the sake of 
simplicity. 
Writing $\ket{\Gamma}$ in Eq.~(\ref{e.Psi0})
on the basis of the $\hat O_\Gamma$-eigenstates, from 
Eq.~(\ref{e.H}) it follows
\begin{equation}
\ket{\Psi(t)}=\sum_{\gamma}c_{\gamma}\ket{\gamma} 
\otimes\ket{\Xi^\gamma(t)}~
\label{e.Psit}
\end{equation}
at any time during the pre-measurement ($0{<}t{<}T$), with
\begin{equation}
\ket{\Xi^\gamma(t)}\equiv
e^{-it\hat{H}^\gamma_\Xi}\ket{\Xi}~,
\label{e.Xigammat}
\end{equation}
and 
\begin{equation}
\hat H^\gamma\equiv u\,\omega_\gamma\hat O_\Xi+\hat H_\Xi~.
\label{e.Hgamma}
\end{equation}
In the standard model, the possibility of extracting 
information about $\Gamma$ reporting on $\Xi$, 
relies on the $\gamma$-dependence of $\ket{\Xi^\gamma(t)}$,
i.e. on the dynamical entanglement generation induced by the coupling  
$u\hat{O}_\Gamma\otimes\hat{O}_\Xi$ if, and only if,
$\ket{\Xi}$ is not a $\hat{O}_\Xi$-eigenstate.
On the other hand, the measuring apparatus is expected to be in a 
stationary state before the above coupling is switched on.
Therefore, it is usually taken
\begin{equation}
\hat H_\Xi\ket{\Xi}=E_0\ket{\Xi}~~~{\rm and}~~~[\hat O_\Xi,\hat 
H_\Xi]\neq 0~.
\label{e.commOH}
\end{equation}

The evolution described by Eq.~(\ref{e.Psit}) 
can be studied by the PRECS\cite{CCGV13osid}, a method
based on the use of generalized coherent states 
\cite{ZhangFG90,Perelomov72,ComberscureD12}  for $\Xi$, whose 
construction, as far as the model (\ref{e.H}) is 
concerned, can be 
summarized as follows.
Consider the operators $\hat{O}_\Xi$ and $\hat{H_\Xi}$:
they will generally be elements of a Lie group $G$, usually dubbed 
(environmental) {\it dynamical group}, and, in most physical situation, 
they also belong to the related Lie algebra {\bf g} (in that they are 
linear combination of the group generators).
The arbitrary choice of a reference 
state $\ket{R}{\in}\mathcal{H}_\Xi$ defines the subgroup $F$ of the 
operators $\hat{f}$ acting trivially on $\ket{R}$,  i.e. such that 
$\hat{f}\ket{R}{=}e^{i\phi_f}\ket{R}$, and hence the
environmental coset $G/F$.
Environmental coherent states (ECS) are the states
\begin{equation}
\ket{\Omega}=\hat{\Omega}\ket{R}~{\rm with}~\hat{\Omega}\in 
G/F~.
\label{e.ECS}
\end{equation}
It is demonstrated\cite{ZhangFG90,Perelomov72,ComberscureD12} that 
coherent states $\ket{\Omega}$ are in one-to-one correspondence with 
points $\Omega$ on a differentiable manifold ${\cal M}$. 
The construction of ECS entails the definition of an invariant (with 
respect to $G$) measure $d\mu(\Omega)$ on ${\cal M}$, as 
well as of a metric tensor ${\bf m}$. Moreover, ECS form an 
over-complete set on ${\cal H}_\Xi$, 
and provide an identity resolution in the form
\begin{equation}
\hat{{\rm I\!I}}_\Xi=\int_{\cal M}\, 
d\mu(\Omega)\ket{\Omega}\bra{\Omega}~,
\label{e.CoherentIdentity}
\end{equation}
where $\hat{\rm I\!I}_\Xi$ is the identity operator on ${\cal H}_\Xi$.
Getting back to the model (\ref{e.H}),
if {\bf g} is semi-simple, referring to its Cartan basis
\footnote{
It is the basis of a semi-simple Lie algebra, usually indicated
by $\{\{\hat H_i\},\{\hat E_\alpha, \hat E_{-\alpha}\}\}$, such that
$[\hat H_i,\hat H_j ]{{=}}0$,
$[\hat H_i,\hat E_\alpha ]{{=}}\alpha_i \hat E_\alpha$,
$[\hat E_\alpha,\hat E_{-\alpha}]{{=}}a_i \hat H_i$, and
$[\hat E_\alpha, \hat E_\beta ]{{=}}c_{\alpha\beta}\hat E_{\alpha+\beta}$.
A Hamiltonian which is linear in the generators is said to be in the
{\it canonical} form when it is
$\hat H{{=}}\sum_i \nu_i \hat H_i{+}\sum_\alpha \epsilon_\alpha \hat E_\alpha+
\epsilon_\alpha^* \hat E^\dagger_\alpha$,
with $\hat E_\alpha^{(\dagger)}$ the shift-up(down) operators.},
and reminding condition (\ref{e.commOH}), one recognizes that $\hat 
H^\gamma$ is in the canonical form with $\hat O_\Xi$ hermitian 
linear combination of shift-up and -down operators
:
This usually entails the choice of $\ket{R}$ 
as the eigenstate of $\hat H_\Xi$ such that $\hat O_\Xi\ket{R}{{=}}0$, 
that naturally provides ${\cal M}$ with a symplectic 
structure~\cite{Onofri75,Simon80}, i.e. ${\Omega}{=}(z,\bar z)$ 
with $z,\bar z$ canonical coordinates, and $d\mu(\Omega){{=}}{\rm 
det}({\bf m})dzd\bar z$.
Consistently with conditions \eqref{e.commOH}, we can set 
$\ket{\Xi}{=}\ket{R}$, implying
\begin{equation}
\exval{\Xi|\hat O_\Xi|\Xi}=0~.
\label{e.exvalO-R}
\end{equation}

Coherent states have peculiar dynamical properties,
which are often summarized by the motto {\it "once
a coherent state, always a coherent state"}\cite{ZhangFG90}.
In the specific case of a system ruled by the Hamiltonian 
(\ref{e.H}), with initial state \eqref{e.Psi0}, 
these properties, complemented with the choice 
$\ket{R}{{=}}\ket{\Xi}$,
allows one to write\cite{CCGV13osid}
\begin{equation}
\ket{\Xi^\gamma(t)}=
e^{i\varphi^\gamma_t}\ket{\Xi^\gamma_t}~,
\label{e.R-gamma-t}
\end{equation}
where $\ket{\Xi^\gamma_t}$ is the coherent state corresponding to the
point ${\Xi}^\gamma(t)$ along the trajectory on ${\cal
M}$ defined by the solution of the classical-like equations of motion
\begin{equation}
i{\rm m}_{z\bar z}\frac{d z}{dt}=\frac{\partial}{\partial\bar z}
H^\gamma(\Omega)~,~{\rm and~the~same~with}~ z\leftrightarrow\bar z~,
\label{e.eom}\
\end{equation}
with $H^\gamma(\Omega){\equiv}\exval{\Omega|\hat H^\gamma|\Omega}$ and 
$\Xi^\gamma(0){{=}}0$;
as for the phase factor it is
$\varphi^\gamma_t{{=}}
\int_0^t~dy~
\exval{\Xi^\gamma_y|\left(i\frac{\partial}{\partial y}-
\hat H^\gamma\right)|\Xi^\gamma_y}$.

Once the ECS are constructed, any state of the composite system $\Psi$ 
can be parametrically represented by formally splitting $\Gamma$ and 
$\Xi$ through the insertion of $\hat{\rm I\!I}_\Xi$ in the form 
(\ref{e.CoherentIdentity}), as 
shown in Ref.~\cite{CCGV13pnas}. In particular, exploiting the fact that 
$d\mu(\Omega)$ is group-invariant, the state (\ref{e.Psit})
reads
\begin{equation}
\ket{\Psi(t)}=\int_{\cal M}\,d\mu(\Omega)
\chi_t(\Omega)~\ket{\phi_t(\Omega)}\otimes \ket{\Omega}~,
\label{e.Psipara}
\end{equation}
with
\begin{equation}
|\phi_t(\Omega)\rangle=\frac{1}{\chi_t(\Omega)}
\sum_\gamma c_\gamma\exval{\Omega|\Xi^\gamma(t)}~\ket{\gamma}~,
\label{e.phipara}
\end{equation}
\begin{equation}
\chi^2_t(\Omega)=
\sum_\gamma|c_\gamma|^2h^\gamma_t(\Omega)~~,~~
h^\gamma_t(\Omega)=|\exval{\Omega|\Xi^\gamma_t}|^2~,
\label{e.chi2-hgammat}
\end{equation}
and we have set $\chi_t(\Omega)$ in $R^+$ by choosing its arbitrary 
phase equal to $0$. 
Equations (\ref{e.Psipara}-\ref{e.chi2-hgammat}) define the 
parametric
representation with ECS of $\ket{\Psi(t)}$.
Notice that the dependence of the principal system's pure 
states $\ket{\phi_t(\Omega)}$ on $\Omega$ is the 
signature that $\Gamma$ and $\Xi$ are entangled\cite{CCGV13pnas}.
Moreover, due to $\exval{\Psi(t)|\Psi(t)}{{=}}1$, it is
$\int_{\cal M} d\mu(\Omega)\chi^2_t(\Omega){{=}}1$ at any time,
which allows one to interpret $\chi^2_t(\Omega)$ as the normalized 
density distribution of ECS on the manifold ${\cal M}$
\cite{ZhangFG90,CCGV13pnas}.
\section{Informative apparatus}
\label{S.Decoherence}
The essential goal of any pre-measurement is that of setting a cogent relation
between elements of $\{\ket{\gamma}\}_{{\cal H}_\Gamma}$ and pointer 
states\cite{Zurek81} of the measuring apparatus. 
Referring to Eq.~(\ref{e.Psit}), this implies that
\begin{eqnarray}
&&\!\!\!\!\!\!\!\!\!\!\!\!\!\!\!\!\!\!\!\!\!\!\!\!\!\!\!\!\!\!\!\!\!
{\rm different~states}~
\ket{\Xi^\gamma(\tau)}~{\rm and}~
\ket{\Xi^{\gamma'}(\tau)}~,\label{e.differentstates}\\
&&\!\!\!\!\!\!\!\!\!\!\!\!\!\!\!\!\!\!\!\!\!\!\!\!\!\!\!\!\!\!\!\!\!
{\rm produce~different~outcomes~whenever}~\gamma\neq\gamma'~,
\label{e.differentoutputs}
\end{eqnarray}
for $\tau{>}\tau_{\rm d}$, where 
$\tau_{\rm d}$ is the time when the pre-measurement can be 
considered successfully concluded.
Condition (\ref{e.differentstates}-\ref{e.differentoutputs}) 
can be translated\cite{L-SCV15ijtp} into some
property that $\chi^2_\tau(\Omega)$ must feature
in order to describe an informative apparatus:
in fact, defining the $\varepsilon$-support of each
component $h^\gamma_t(\Omega)$ as the region ${\cal S}^\gamma_t{\in} {\cal 
M}$ such that $h^\gamma_t(\Omega){>}\varepsilon~\forall \Omega{\in} 
{\cal S}^\gamma_t$, with $\varepsilon$ a small number in $R^+$, the 
request (\ref{e.differentstates}-\ref{e.differentoutputs}) can be 
identified\cite{L-SCV15ijtp} with the condition
\begin{equation} 
{\cal{S}}^\gamma_\tau\cap{\cal{S}}^{\gamma'}_\tau=\emptyset~~
{\rm for}~~\gamma\neq\gamma'~.
\label{e.distinguishability} 
\end{equation} 

\begin{figure}[h]
\includegraphics[width=80mm,angle=0]{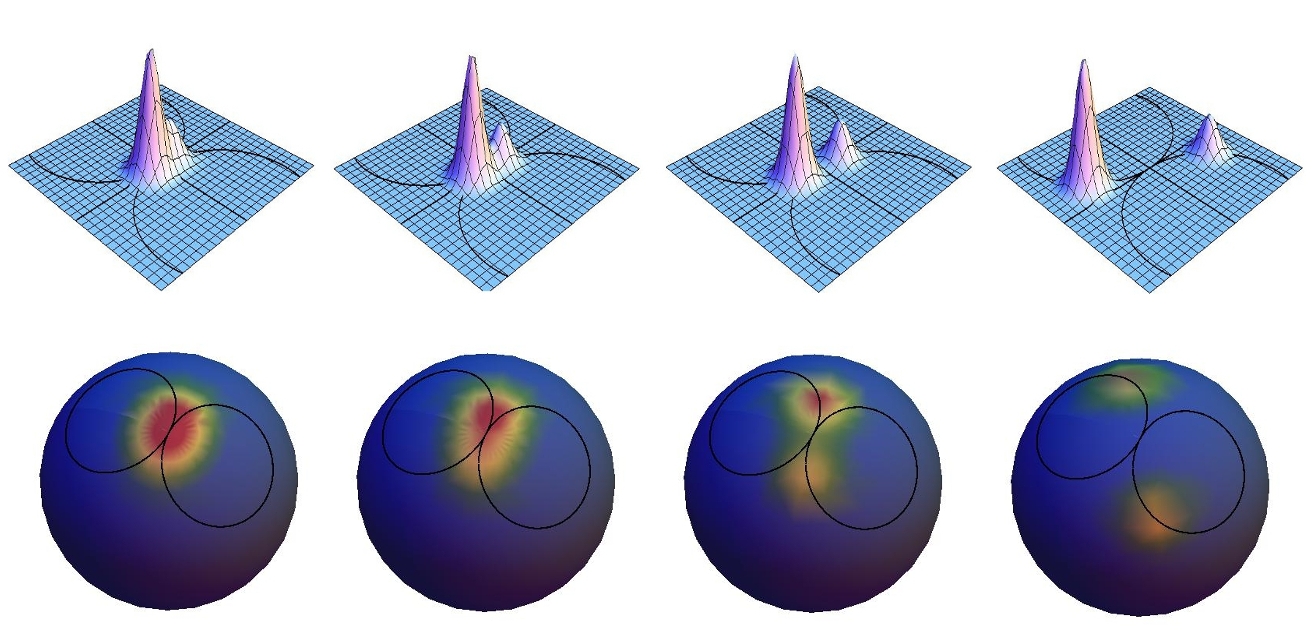}
\caption{Evolution of $\bar\chi^2_t(\Omega)$ for the 
models 
\eqref{e.qubit-boson} and \eqref{e.qubit-spinJ}, top and bottom panels, 
respectively, as time goes by from left to right.
Black lines indicate the trajectories $\Xi^\gamma_t$, with $\gamma{=}\pm$.}
\label{f.1}
\end{figure}

\begin{figure}[b!]
\includegraphics[width=80mm,angle=0]{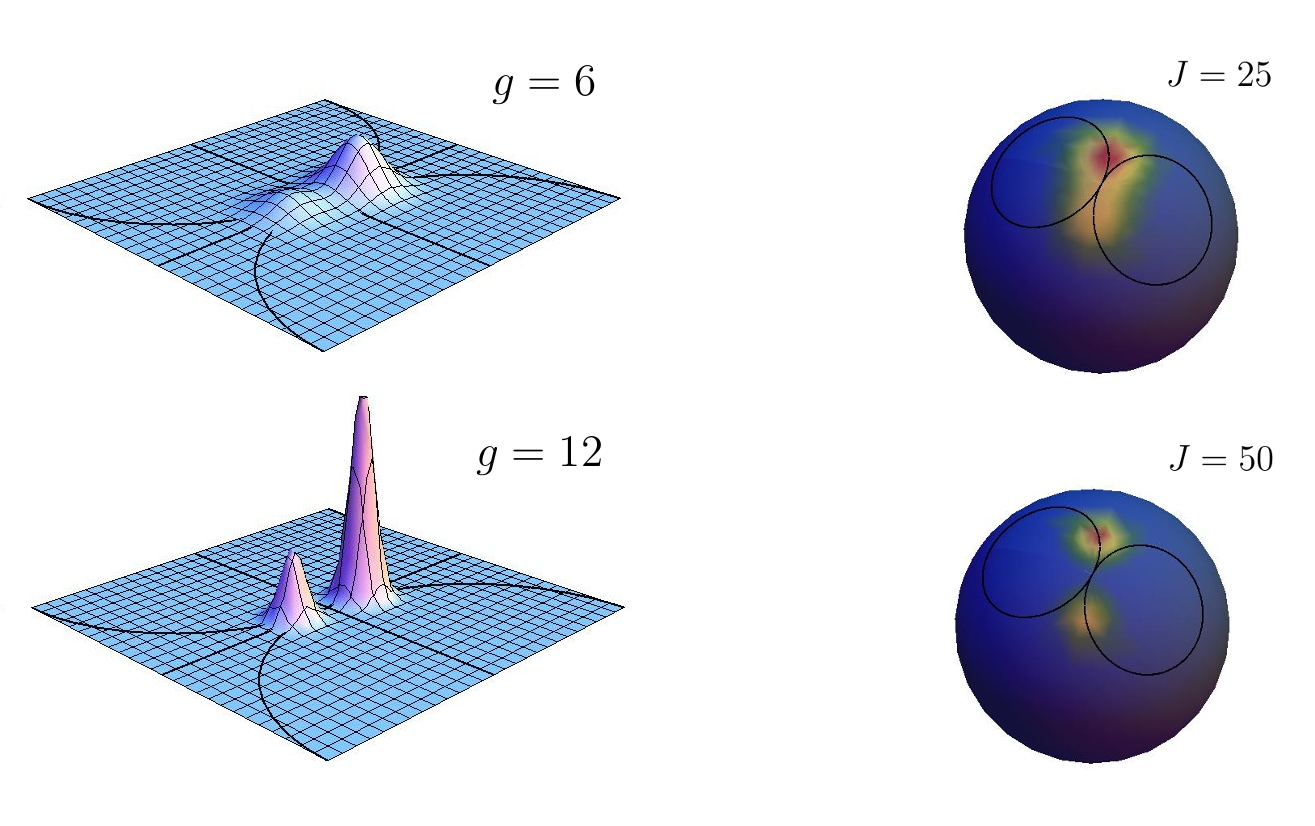}
\caption{$\bar\chi^2_t(\Omega)$ at a given time for 
the 
models
\eqref{e.qubit-boson} and \eqref{e.qubit-spinJ} 
with different parameters $g$ and $J$ - left and right panels 
respectively - Black lines as in Fig.1.}
\label{f.2}
\end{figure}

The one-to-one correspondence thus established between 
each function $h^\gamma_\tau$ and the region ${\cal 
S}^\gamma_\tau$ in 
${\cal M}$
allows one to write  
\begin{equation} 
\ket{\Psi(\tau)}=\sum_\gamma c_\gamma\!\int_{{\cal S}^\gamma_\tau}\,
d\mu(\Omega)e^{i\varphi^\gamma_t}\sqrt{h^\gamma_\tau (\Omega)}
~\,\ket{\gamma}\otimes\ket{\Omega}~
\label{e.psiparadeco} 
\end{equation}
for all $\tau$ in the time interval where 
condition (\ref{e.distinguishability}) holds.
Notice that, despite such condition
specifically concern the apparatus, Eq.~(\ref{e.psiparadeco}) strictly 
implies that 
decoherence with respect to the basis $\{\ket{\gamma}\}_{{\cal 
H}_\Gamma}$ has occurred\cite{Liuzzo-ScorpoMaster14,L-SCV15ijtp},  and 
the time 
$\tau_{\rm d}$ is consistently recognized as a decoherence time for $\Gamma$.
For the sake of a lighter notation we will hereafter drop the $\tau$ 
index in ${\cal S}^\gamma_\tau$ understanding that, whenever ${\cal 
S}^\gamma$ is referred to, it is $\tau{>}\tau_{\rm d}$.
For a more transparent discussion, it is now worth considering two 
models with Hamiltonian of the form 
$(\ref{e.H})$, that have been introduced
as paradigmatic ones for describing the decoherence 
phenomenon, and extensively studied in various 
contexts~\cite{Schlosshauer07,LeggetEtal87}.
They are described by
\begin{equation}
\hat{H}_{qb}=\nu \hat{b}^\dagger\hat{b} 
+g\hat{\sigma}^z\otimes(\hat{b}+\hat{b}^\dagger)~,
\label{e.qubit-boson}
\end{equation}
and
\begin{equation}
\hat{H}_{qS}=h\hat{J}^z+\mu\hat{\sigma}^z\otimes\hat{J}^x
\label{e.qubit-spinJ}~,
\end{equation}
where $\hat{\sigma}^z$ is the $Z$-Pauli matrix, $\hat{b}$ and 
$\hat{b}^\dagger$ are bosonic operators 
($[\hat{b},\hat{b}^\dagger]{=}1$), and $\hat{J}^\pm{=}\hat{J}^x\pm 
i\hat{J}^y$ are spin-$J$ operators 
($[\hat{J}^\alpha,\hat{J}^\beta]{=}
i\varepsilon^{\alpha\beta\gamma}\hat{J}^\gamma$).
The respective ECS are those usually referred to as field and spin 
coherent states, with $\ket{R}$ such that $\hat{b}\ket{R}{=}0$ and 
$\hat{J}^-\ket{R}{=}0$, and manifold $\cal M$ the complex plane and the 
unit $2$-sphere.
The trajectories ${\Xi}^\gamma_t$ defining the states 
$\ket{\Xi^\gamma_t}$ in 
Eqs.~(\ref{e.phipara}-\ref{e.chi2-hgammat}),  with $\gamma{=}\pm$, can be 
explicitly 
determined~\cite{Liuzzo-ScorpoMaster14} and are shown in Fig.\ref{f.1} as 
black lines on the respective manifold. 
In the same figure, the distributions 
$\bar\chi^2_t(\Omega){\equiv}{\rm det}({\bf m})\chi^2_t(\Omega)$ are
plotted at various times: it strikes that,
as time goes by, they  acquire a multi-modal structure, 
with as many distinct modes as the number of 
different $\gamma$ in the $\hat{O}_\Gamma$-spectrum, thus visualizing 
how condition \eqref{e.distinguishability} is dynamically achieved.
Moreover, considering $\bar\chi^2_t(\Omega)$ for different values 
of $g$ and $J$, as in Fig.~\ref{f.2}, reveals that when
such parameters increase, each $\gamma$-peak becomes more pronounced, 
and the corresponding support ${\cal{S}^\gamma}$ consequently 
shrinks around ${\Xi}^\gamma_t$.
This evidence, that reflects the more general 
result\cite{Yaffe82} briefly stated at point {\it ii)} of the next 
section, is clearly reminiscent of some sort of 
classical limit for $\Xi$. 
The description of the process which is now 
available 
allows us to push the formalism towards a well 
defined macroscopic limit for the measuring apparatus only, that leave 
the principal system unaffected.

\section{Large-N limit}
\label{S.Large-N limit}

This Section is based on a work by L.G.Yaffe\cite{Yaffe82}, dealing with the 
fundamental question ``\emph{Can one find a classical system whose 
dynamics is equivalent to some $N\rightarrow\infty$ limit of a given 
quantum theory?}'', where $N$ is some measure of the number of dynamical 
variables. An extensive discussion of Ref.~\cite{Yaffe82} goes beyond 
the scope of this letter, but briefly retracing the reasoning underlying 
the results which are most relevant to get to our final goal, is 
necessary. In doing that we will try to keep contact with what we have 
described so far.

\noindent Given that:

- a classical theory, $C$, is defined by a phase space $\cal C$, a 
Poisson bracket, and a classical Hamiltonian $H_{\rm cl}$;

- a quantum theory, $Q$, is defined by a Hilbert space, a Lie 
algebra, and a Hamiltonian $\hat{H}$ (and dynamical group $G$),

\noindent then:

{\it i)} Be $Q_\kappa$ a quantum theory
characterized by some (quanticity) parameter $\kappa$, assumed to take 
positive real values including the limiting $\kappa{=}0$ one.
This is the theory that describes the apparatus $\Xi$ in the above 
sections, in particular, $\kappa{=}1/g$ or $1/J$ in the models 
(\ref{e.qubit-boson}) or (\ref{e.qubit-spinJ}), respectively. 

{\it ii)} It exists a minimal set of conditions that $Q_\kappa$ must 
fulfill to guarantee that its $\kappa\to 0$ limit is a classical theory 
$C$. Such conditions emerge in terms of coherent states 
for the dynamical group of the theory, $G_\kappa$, 
and establish a one-to-one correspondence, ${\cal M}_\kappa\to{\cal C}$,
between points on the related manifold and on the phase-space of $C$. 
These coherent states are the ECS defined in the previous section, and 
one of the above conditions implies
\begin{equation}
h^\gamma_t(\Omega){\rm det}({\bf m})~~\underset{\kappa\to 0}{\longrightarrow}~~
\delta(\Omega-\Xi^\gamma_t)~,
\label{e.hdelta}
\end{equation}
as suggested in Fig.~2.

{\it iii)} Be $Q_N$ a quantum many-body (field) theory with some 
global $X(N)$ symmetry ($X{=}O, Sp, U $...), dynamical group $G_N$, 
and related manifold ${\cal M}_N$.
This is the microscopic quantum theory that would exactly describe the 
apparatus $\Xi$, were we able to determine the details of its internal 
interactions as well as of those between each one of its components and 
$\Gamma$.

{\it iv)} Any such $Q_N$ theory defines a $Q_\kappa$ one, by this 
meaning that the latter can be explicitly defined from the former, such 
that $\kappa{=}1/N$ (or $1/N^2$, depending on specific features of $Q_N$). 
The relation between the two theories is established via their 
respective dynamical groups, $G_N$ and $G_\kappa$, by making them be 
different representations of the same algebra. Operators 
$\hat{A}_\kappa$ and $\hat{A}_N$ in the two different theories are also 
formally related (though it is not possible to explicitly express such 
relation in general). 

{\it v)} 
It is demonstrated that $\lim_{N\to\infty} Q_N{=}\lim_{\kappa\to 
0}Q_\kappa{=}C$, where the last equality means that $Q_\kappa$ fulfills the 
conditions of point {\it ii)} as $\kappa\to 0$. The above chain of 
equalities implies the existence of a mapping from ${\cal M}_N\to 
{\cal M}_\kappa\to{\cal{C}}$, with the last correspondence
biunivocal, as from point {\it ii)}.

Resulting from the above is the following:
to each $\ket{\Omega}$, coherent state for $Q_\kappa$, it is associated a set 
$\{\ket{\Omega\scriptscriptstyle{N}}\}_\sim$ of coherent 
states for $Q_N$ such that
\begin{equation}
\lim_{N\to\infty}\exval{\Omega{\scriptscriptstyle{N}}_i|\hat{A}_N|\Omega{\scriptscriptstyle{N}}_i}=
\lim_{k\to 0}\exval{\Omega|\hat{A}_\kappa|\Omega}
{\equiv} A_{\rm cl}(\Omega)~,
\label{e.classicallyequivalent}
\end{equation}
for all $\ket{\Omega\scriptscriptstyle{N}_i}{\in}
\{\ket{\Omega{\scriptscriptstyle{N}}}\}_\sim$ and any 
hermitian operator $\hat A_\kappa$ with $A_{\rm cl}(\Omega){<}\infty$. 
Elements of the same set are related 
by $\ket{\Omega{\scriptscriptstyle{N}}_i}{=}
{\cal U}_{ij}\ket{\Omega{\scriptscriptstyle{N}}_j}$ with
${\cal U}_{ij}$ any unitary $X(N)$-symmetry operation.
Operators $\hat{A}_\kappa$ and $\hat{A}_N$ are related as from point 
{\it iv)}, and $A_{\rm cl}$ is a real function on $\cal C$, with $A_{\rm 
cl}(\Omega)$ its value on the point that univocally corresponds to 
$\ket{\Omega}$, according to point {\it ii)}.
Verbalizing Eq.~(\ref{e.classicallyequivalent}), states in
$\{\ket{\Omega{\scriptscriptstyle{N}}}\}_\sim~$ are dubbed {\it classically 
equivalent}, and operators such that  their symbols 
$\exval{\Omega{\scriptscriptstyle{N}}|\cdot
|\Omega{\scriptscriptstyle{N}}}$, keep finite 
for $N\to\infty$ are called {\it classical operators}\cite{Yaffe82}. 
\begin{center}
\begin{figure}[h]
\includegraphics[width=80mm,angle=0]{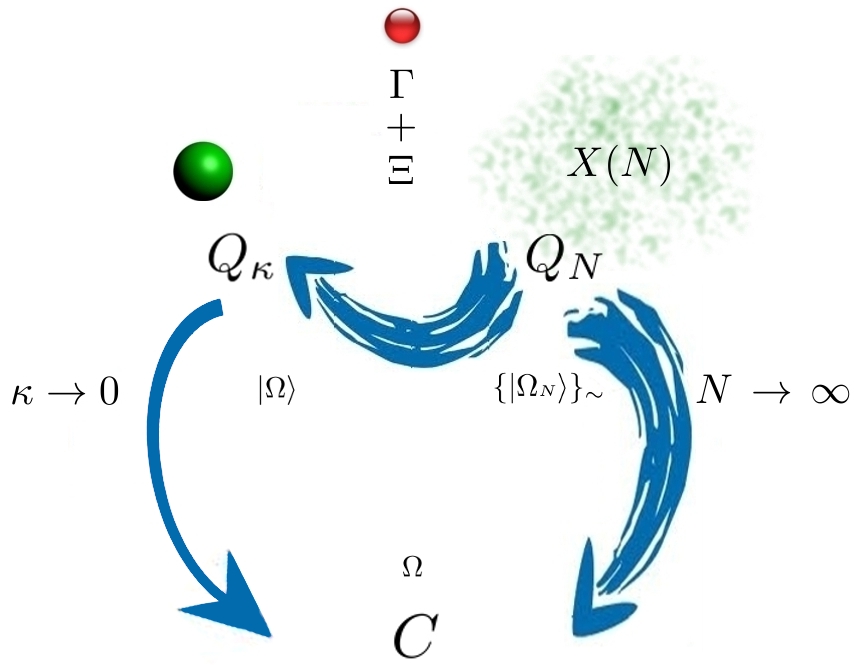}
\caption{Graphical depiction of the relation between effective and 
microscopic theories, $Q_\kappa$ and $Q_N$, giving rise to the same 
classical theory $C$.}
\label{f.boson-t}
\end{figure}
\end{center}
\section{Macroscopic measuring apparatus}
Regarding the measurement process, Yaffe's results teaches 
us that the classical limit ($\kappa\to 0$) of the effective theory 
$Q_\kappa$ used for describing $\Xi$ during the 
pre-measurement, can keep describing a macroscopic 
($N\to\infty$) apparatus; however, in order for this to be the case, an 
$X(N)$-invariant $Q_N$ theory must actually underlie $Q_\kappa$, being 
the one that would provide us with the exact, microscopic description of 
$\Xi$, if we were able to deal with it in the large-$N$ limit.

We can now get back to the PRECS treatment, to find that the theory 
$Q_\kappa$, with its related coherent states $\ket{\Omega}$, is already 
well defined, and  the symbols $\exval{\Omega|\hat A_k|\Omega}$ 
are the Husimi functions\cite{ZhangFG90,Perelomov72,ComberscureD12} of 
operators acting 
on ${\cal H}_\Xi$. Therefore, from condition (\ref{e.distinguishability}) and 
Eq.~\eqref{e.classicallyequivalent}, we find that to each
$\cal{S}^\gamma$, and hence to each trajectory in ${\cal M}$, it 
corresponds a set  
$\{\ket{\Omega{\scriptscriptstyle{N}}}\}^\gamma_\sim$ of classically 
equivalent coherent states of the microscopic, $X(N)$-invariant, 
quantum theory $Q_N$.

States belonging to the same set 
$\{\ket{\Omega{\scriptscriptstyle{N}}}\}^\gamma_\sim$
keep being distinct in the large-$N$ limit, as this limit does not 
affect the transformations ${\cal U}_{ij}$ that relate them. Therefore, 
the set ${\cal V}^\gamma_N$ of points corresponding to 
$\{\ket{\Omega{\scriptscriptstyle{N}}}\}^\gamma_\sim$ 
in ${\cal M}_N$ has a finite volume $V^\gamma_N$ even if the apparatus 
becomes macroscopic.

States belonging to sets 
$\{\ket{\Omega{\scriptscriptstyle{N}}}\}^\gamma_\sim$ 
labeled by different $\gamma$s are related by
\begin{eqnarray}
&\!&\ket{\Omega{\scriptscriptstyle{N}_i}(t)}^{\!\gamma'}=
U^{\gamma\gamma'}_N(t)
\ket{\Omega{\scriptscriptstyle{N}_i}(t)}^{\!\gamma}
\label{e.gammaevolutions}
\\
\!\!\!\!\!\!\!\!\!\!\!\!\!\!{\rm with ~~~}&\!&U^{\gamma\gamma'}_N(t)\equiv
e^{-it\hat H^{\gamma'}_N+it\hat H^{\gamma}_N}~,
\label{e.Ugammagamma'}
\end{eqnarray}
which follows from the fact that evolutions defined by different 
$\gamma$s have the same initial state.
For $t{{>}}\tau_{\rm d}$ they are not classically equivalent, implying 
${\cal V}^\gamma_N\cap{\cal V}^{\gamma'}_N{=}\emptyset$,
but yet they have the same energy in the $N\to\infty$ limit,
\begin{eqnarray}
&\!&\!\!\!\!\!\!\!\!\!\!\!\!\!\!\!\!
\lim_{N\to\infty}\exval{\Omega{\scriptscriptstyle{N}}|\hat 
H^\gamma_N|\Omega{\scriptscriptstyle{N}}}
\underset{\Omega{\scriptscriptstyle{N}}\in\cup_\gamma{\cal V}^\gamma_N}{=}
\label{e.degN}\\
&\!&\!\!\!\!\!\!\!\!\!\!\!\!\!\!\!\!
\lim_{\kappa\to 0}\exval{\Omega|\hat H^\gamma|\Omega}
\underset{\Omega\in\cup_\gamma {\cal S}^\gamma}{=}
H_{\rm cl}(\Xi^\gamma_\tau)=E_0~,
\label{e.degQ}
\end{eqnarray}
as seen from Eqs.~
\eqref{e.classicallyequivalent}, \eqref{e.hdelta}, \eqref{e.exvalO-R}, 
and \eqref{e.commOH}.

The above analysis tells us that if one were to study the 
behaviour of a macroscopic measuring apparatus in terms of its 
microscopic quantum theory, despite not being able to do it exactly 
due to the large number of dynamical variables, yet she could 
extract information on $\Xi$, for $t{{>}}\tau_{\rm d}$ and $N\to\infty$, from 
the degenerate coherent states
$\cup_\gamma\{\ket{\Omega{\scriptscriptstyle N}}\}^\gamma_\sim$, grouped 
into disjoint sets of classically equivalent ones.

\section{Outcome production} 

Suppose now that, at a certain time $T{>}\tau_{\rm d}$, a local 
perturbation acts on some parts of $\Xi$, thus breaking the global 
$X(N)$-symmetry. The relation between $Q_N$ and $Q_\kappa$ is 
consequently broken, and the latter theory cannot be further used for 
describing the apparatus. The only tool with which we are left for 
studying $\Xi$ is the  microscopic theory $Q_N$, and the information 
collected upon it in the previous section.
According to the usual description of symmetry breaking (SB) in systems 
made by a large number of particles, and exclusively focusing upon the 
$N\to\infty$ limit
of $Q_N$, the above $X(N)$-SB will select some states amongst those that 
equivalently describe the measuring apparatus immediately before $T$, 
by making their energy lower than that of all the others. 
Specifically, as the operator $\hat\delta$ representing the 
local SB-perturbation cannot commute with any of the 
global transformations $\hat{U}^{\gamma\gamma'}_N$ in 
Eqs.~(\ref{e.gammaevolutions}-\ref{e.Ugammagamma'}), neither can it
depend on the spectrum $\{\omega_\gamma\}$, it is
$^{\gamma}\!\exval{\Omega{\scriptscriptstyle{N}_i}|\,\hat\delta\,|
\Omega{\scriptscriptstyle{N}_i}}^{\!\gamma}{\neq}
^{\gamma'}\!\!\exval{\Omega{\scriptscriptstyle{N}_i}|\,\hat\delta\,|
\Omega{\scriptscriptstyle{N}_i}}^{\!\gamma'}$ 
unless they both vanish.
In other terms, a local perturbation on $\Xi$ that be independent on the 
previous evolution of $\Psi$, cannot cause the same energy-lowering in 
states belonging to different sets of classically equivalent coherent states.

Therefore, only one $\gamma_{\rm out}$ will be selected
\begin{equation}
\cup_\gamma\{\ket{\Omega{\scriptscriptstyle{N}}}\}^{\gamma}_\sim~~~
\underset{X(N){\text{-}{\rm SB}}}{-\!\!-\!\!-\!\!-\!\!-\!\!\longrightarrow}~~~
\{\ket{\Omega{\scriptscriptstyle{N}}}\}^{\gamma_{\rm out}}_\sim~.
\label{e.objectification}
\end{equation}
Picking one specific set of classically equivalent coherent 
states, i.e. one specific $\gamma_{\rm out}$, ensures that all
classical functions get their respective definite value $A_{\rm 
cl}(\Xi^{\gamma_{\rm out}}_T)$ on the classical phase-space ${\cal 
C}$ via Eq.~\eqref{e.classicallyequivalent},
and a classical behaviour unambiguously emerges for the 
measuring apparatus. In particular, the classical function 
corresponding to $\hat O_\Xi$ 
will take the value $O_{\rm cl}(\Xi^{\gamma_{\rm out}}_T)$, which will 
be the result of the measurement process.

Notice that the effective theory describing the apparatus during the 
pre-measurement is $X(N)$-symmetric if and only if the interaction between 
$\Gamma$ and $\Xi$ also features such symmetry; therefore, the
$X(N)$-SB is only made possible by the outwards opening of $\Psi$, i.e. 
by enlarging the system considered during the pre-measurement
from $\Psi{=}\Gamma+\Xi$  to $W{=}\Psi+R$, where by $R$ we mean 
the "rest of the world". 
The action of $R$ on $\Xi$ can be controlled, just like in an 
actual measurement where it is triggered by the 
reading of the apparatus. It can also be completely random, in which 
case the $X(N)$-SB induces the emergence of classicality.
Whatever the situation, such action is uncorrelated with the 
dynamics of $\Psi$ before the symmetry breaking.

\section{Born's rule} 

In order to understand {\em what} value of $\gamma_{\rm out}$ 
one should in principle expect, let us go back to our 
measuring apparatus after the pre-measurement is concluded but 
before the SB has occurred ($\tau_{\rm d}{<}\tau{<}T$):
In previous sections we have learned that its macroscopic ($N\to\infty$) 
behaviour follows from the features of the ensemble 
$\{\{\ket{\Omega{\scriptscriptstyle{N}}}\}^{\gamma_1}_\sim,
   \{\ket{\Omega{\scriptscriptstyle{N}}}\}^{\gamma_2}_\sim,...\}$ 
of degenerate coherent states, grouped into disjoint sets, each labeled 
by a specific $\gamma$. States belonging to the same set correspond, as 
$N\to\infty$ and at any time $\tau{>}\tau_{\rm d}$, to the same point 
$\Xi^\gamma_\tau$ in ${\cal C}$, and hence to the same possible outcome
$O_{\rm cl}(\Xi^{\gamma}_T)$. On the other hand, 
due to degeneracy and to the fact that the perturbation causing the 
$X(N)$-SB is uncorrelated with the evolution of the overall system $\Psi$,
states belonging to the above ensemble are all equally likely.
Therefore, the principles of statistical mechanics tell us that
the probability $p(\gamma_{\rm out})$ of the outcome
$O_{\rm cl}(\Xi^{\gamma_{\rm out}}_T)$ is proportional to the volume 
$V^\gamma_N$ occupied by the
set ${\cal V}^{\gamma_{\rm out}}_N$ of representative points in 
${\cal M}_N$, at 
$\tau{=}T$ and as $N\to\infty$.
Being left with the final problem of evaluating $V^\gamma_N$, we 
consider the following:
prior to the symmetry breaking, the two theories $Q_N$ and $Q_\kappa$, 
in their respective large-$N$ and $\kappa\to 0$ limit, describe the same 
classical behaviour of $\Xi$. Therefore, getting back to
$\Psi{=}\Gamma+\Xi$, for $\tau{>}\tau_{\rm d}$ it is
\begin{equation}
\exval{\Psi|{\mathbb{I}}_{\cal{H}_\gamma}\otimes\hat{A}|\Psi}=
\sum_\gamma\int_{\cal{S}_\gamma} 
d\mu(\Omega)\chi^2_\tau(\Omega)\exval{\Omega|\hat{A}|\Omega}
\end{equation}
and it must be
\begin{equation}
\lim_{\kappa\to 0}
\exval{\Psi|{\mathbb{I}}_{\cal{H}_\gamma}\otimes\hat{A}|\Psi}=
\lim_{N\to\infty}\sum_\gamma\int_{{\cal V}^\gamma_N}
d\mu(\Omega{\scriptscriptstyle{N}})
\exval{\Omega{\scriptscriptstyle{N}}|\hat{A}_N|\Omega{\scriptscriptstyle{N}}}
\label{e.samexpectationvalue}
\end{equation}
for any {\em classical} operator $\hat A$ acting on  ${\cal H}_\Xi$, and
for all initial states $\ket{\Gamma}$ (i.e. for 
all possible sets of $\{c_\gamma\}$ such that 
$\sum_\gamma|c_\gamma|^2{=}1$).
Given Eq.~(\ref{e.classicallyequivalent}) this is seen to require
\begin{eqnarray}
V^\gamma_N\propto&\lim_{N\to\infty}&
\int_{{\cal V}^\gamma}d\mu(\Omega{\scriptscriptstyle{N}})\\
=&\lim_{\kappa\to 0}&
\int_{\cal{S}^\gamma}d\mu(\Omega)\chi^2_\tau(\Omega){=}|c_\gamma|^2~,
\label{e.Born}
\end{eqnarray}
which results in the Born's rule
\begin{equation}
p(\gamma_{\rm out})=|c_{\gamma^{\rm out}}|^2~.
\end{equation}

Finally notice that the selection entailed 
by Eq.~\eqref{e.objectification} trails behind itself that of the state 
$\ket{\gamma_{\rm out}}$ for $\Gamma$, as seen from 
Eqs.~(\ref{e.phipara}) and \eqref{e.hdelta},
thus realizing the reduction of the observed system's quantum 
state.

Before moving towards some concluding comments, we note that 
considering degenerate, possibly continuous, sharp observables and/or 
non trivial pointer functions is just a matter 
of a more complicated notation. On the other hand, the generalization to 
un-sharp observables, i.e. POVM rather than projective measurements, 
is a more delicate issue, that will be possibly tackled in future 
works.

\section{Conclusions}

Let us briefly retrace the route that brought us from the initial 
separable state $\ket{\Gamma}\otimes\ket{\Xi}$ to the production of an 
outcome, and the Born's rule. 

We have considered projective measures and, referring to the 
standard model of unitary pre-measurements, analyzed their
dynamics by means of some specific tools, namely generalized coherent 
states and the PRECS. After having expressed the necessity of decoherence 
as a condition on the distribution of quantum states relative to the 
apparatus, we have formally substantiated the intuitive relation between 
\begin{description}
\item[(1)]{ 
the classical limit of the effective quantum theory used in the standard 
model for describing the apparatus by a limited  number of dynamical 
variables, and} 
\item [(2)]{ 
the large-$N$ limit of the microscopic theory 
that would exactly describe, were we able to handle it, a macroscopic 
quantum measuring apparatus.}
\end{description}
This has enlightened that a global symmetry must characterize {\bf (2)} 
if {\bf (1)} is to be given a physical meaning: In fact, the 
breaking of such symmetry
causes the breakdown of the effective quantum description of 
the apparatus adopted during the pre-measurement and 
contextually selects an objective outcome. The fact that a huge number 
of microscopic configurations of the apparatus correspond to 
the same outcome brings probabilities into play, according to the
principles of statistical mechanics and with no consequences on the 
deterministic nature of the quantum description.
In other terms, observing a quantum system with a macroscopic apparatus 
produces a probabilistic result not due to the system being quantum but 
rather because the apparatus is big.

The reader will recognize, all along the description proposed in this 
work, many concepts and formal elements characterizing previous 
approaches to the quantum measurement process, from einselection to 
time-reversal symmetry breaking, from non-linear dynamics of the output 
production to generalized coherent states. 
In fact, we believe that this work does not collide with 
(almost) any of the previously developed analysis of the quantum 
measurement process, but rather seems to reconcile them.

Finally, our description can be tested by artificially causing the 
global-symmetry breaking, after decoherence has occurred in a
pre-measurement process. To this respect, we are investigating the 
possibility of an experimental realization of the model 
\eqref{e.qubit-spinJ} by a spin star, with the external ring playing the 
role of the apparatus, and the global symmetry being the one that ensures its 
total spin $J$ to be fixed during the pre-measurement. 
The possibility of lowering $1/J$ by increasing 
the number of spins on the ring should give us control upon 
the classical and the large-$N$ limits of the apparatus.
Moreover, an action specifically targeted to one single spin of the ring 
should allow us to cause the symmetry breaking and verify that its 
effects are indeed the same as those expected after a quantum 
measurement.

\acknowledgments
We thank F.~Bonechi, D.~Calvani, and M.~Tarlini for their invaluable help.
This work is done in the framework of the
{\em Convenzione operativa} between the Institute for Complex Systems 
of the Italian National Research Council, and the Physics and
Astronomy Department of the University of Florence.

\end{document}